# Simulation of an SEIR infectious disease model on the dynamic contact network of conference attendees

Juliette Stehlé[1], Nicolas Voirin[2,3§], Alain Barrat[1,4], Ciro Cattuto[4], Vittoria Colizza[5,6,7], Lorenzo Isella[4], Corinne Régis[3], Jean-François Pinton[8], Nagham Khanafer[2,3], Wouter Van den Broeck[4] and Philippe Vanhems[2,3]

[1]Centre de Physique Théorique de Marseille, CNRS UMR 6207, Marseille, France

[2]Hospices Civils de Lyon, Hôpital Edouard Herriot, Service d'Hygiène, Epidémiologie et Prévention, Lyon, France

[3]Université de Lyon; université Lyon 1; CNRS UMR 5558, laboratoire de Biométrie et de Biologie Evolutive, Equipe Epidémiologie et Santé Publique, Lyon, France

[4]Complex Networks and Systems Group, Institute for Scientific Interchange (ISI) Foundation, Torino, Italy

[5] INSERM, U707, Paris F-75012, France

[6] UPMC Université Paris 06, Faculté de Médecine Pierre et Marie Curie, UMR S 707, Paris F75012, France

[7]Computational Epidemiology Laboratory, Institute for Scientific Interchange (ISI) Foundation, Torino, Italy

[8]Laboratoire de Physique de l'Ecole Normale Supérieure de Lyon, CNRS UMR 5672, Lyon, France

[§]Corresponding author

Email addresses:

    JS: juliette.stehle@gmail.com

    NV: nicolas.voirin@chu-lyon.fr

    AB: alain.barrat@cpt.univ-mrs.fr

    CC: ciro.cattuto@isi.it

    VC: vcolizza@inserm.fr

    LI: lorenzo.isella@isi.it

    CR: corinneregis@yahoo.fr

    JFP: jfpinton@gmail.com

    NK : nagham.khanafer@chu-lyon.fr

    WVdB : wouter.vandenbroeck@isi.it

    PV : philippe.vanhems@chu-lyon.fr

# Abstract

**Background**

The spread of infectious diseases crucially depends on the pattern of contacts among individuals. Knowledge of these patterns is thus essential to inform models and computational efforts. Few empirical studies are however available that provide estimates of the number and duration of contacts among social groups. Moreover, their space and time resolution are limited, so that data is not explicit at the person-to-person level, and the dynamical aspect of the contacts is disregarded. Here, we want to assess the role of data-driven dynamic contact patterns among individuals, and in particular of their temporal aspects, in shaping the spread of a simulated epidemic in the population.

**Methods**

We consider high resolution data of face-to-face interactions between the attendees of a conference, obtained from the deployment of an infrastructure based on Radio Frequency Identification (RFID) devices that assess mutual face-to-face proximity. The spread of epidemics along these interactions is simulated through an SEIR model, using both the dynamical network of contacts defined by the collected data, and two aggregated versions of such network, in order to assess the role of the data temporal aspects.

**Results**

We show that, on the timescales considered, an aggregated network taking into account the daily duration of contacts is a good approximation to the full resolution network, whereas a homogeneous representation which retains only the topology of the contact network fails in reproducing the size of the epidemic.

**Conclusions**

These results have important implications in understanding the level of detail needed to correctly inform computational models for the study and management of real epidemics.

# Background

The pattern of contacts among individuals is a crucial determinant for the spread of infectious diseases in a population (*1*). The topological structure, the presence of individuals with a much larger number of contacts than the average value (*2-5*), the clustering and presence of well-identified communities of individuals (*6-10*), the frequency and duration of contacts (*11-13*) have important implications for the spread and control of epidemics. Knowledge of contact patterns is critical for building and informing computational models of infectious diseases transmission (*14-23*). Though some of the properties of contact patterns are found to dramatically affect the model predictions (*3-5*), little is known on their empirical characteristics, and few experiments have been conducted to collect data on how individuals mix and interact.

The starting point of most modeling approaches consists in the homogeneous mixing assumption for which every individual has an equal probability of contacting other individuals in the population (*1*). No heterogeneity in the mixing pattern or in the duration or frequency of the contact is considered, and the dynamical nature of the contacts is disregarded. Going beyond this approximation, various approaches have proposed to estimate mixing properties among classes of individuals (e.g. social or age classes) using indirect (*1*) and, more recently, direct (*11, 24-27*) methods. Indirect methods are based on estimating the elements of a "Who Acquires Infection From Whom" (WAIFW) matrix using observed seroprevalence data. In direct methods, each element of a contact matrix is estimated independently of the epidemiologic data. Direct methods rely on data collection about at-risk events via diaries (*11, 12*) or time-use data (*2, 27*). To date, research on human social interaction has been mainly based on self-reported data. Despite a real improvement of the description of potential contacts with respect to a homogeneous mixing approach, self-reported methods involve a limited number of people who provide information on a limited number of snapshots in time (usually one day). The obtained data may suffer from uncontrolled bias and a lack of representativeness since it is not based on objective reports, and the data collection is performed on a random day and generally lacks the longitudinal aspect. These limitations become particularly relevant in the case of contact patterns and infectious diseases transmitted by the respiratory or close-contact route. Here all types of social encounters, even random contacts of very short duration e.g. on public transport, may be important for the transmission, but are rather difficult to report objectively and exhaustively through a diary method.

New technologies are now available that allow the tracking of proximity and interactions of individuals (*28-37*), deeply transforming our ability to understand and characterize social behavior (*38*). Detection of contact patterns can rely on objective and unsupervised measures of proximity behavior, that can be extended to a large number of individuals, with a high temporal and spatial resolution (*28, 30*), thus overcoming the limitations of self-reported data. Departing from the typical static representation of a network of contacts among individuals (*39*), it is now possible to describe the dynamic nature of the interactions. The analysis of the dynamics of a contact network needs to incorporate two essential features: (i) the variations in the duration and frequency of the contacts between individuals, and (ii) the existence of causality constraints in the possible chains of transmission.

Finally, little is known on the level of detail that should be incorporated in the modeling effort to perform in practice realistic simulations of epidemics spreading in a population. Very coarse descriptions of human behavior, such as the homogeneous mixing hypothesis, leave out crucial elements. Conversely, extremely detailed information may yield a lack of transparency in the models, making it difficult to discriminate the impact of any particular modeling assumption or component.

The aim of this paper is to assess the role of the temporal aspects, and of the heterogeneities and constraints of dynamical contact patterns in shaping the dynamics of an infectious disease in a population using data collected during a two-day medical conference. To this aim, we capitalize on the recent development of a data collection infrastructure that allows the tracking of face-to-face proximity of individuals at a high temporal resolution (*28, 30*). We use the data collected during a scientific conference providing the temporal information on individual contact events. This can be mapped into a dynamic network of contacts, where all information on interactions between pairs of individuals, time of occurrence and duration are explicit in the network representation. Along with the explicit dynamic network of contacts, we consider two different projections of the data, defining two types of daily networks that aggregate the empirical data in different ways, reflecting different amounts of available knowledge about the contacts between individuals. We then simulate the spread of an infectious disease over these networks, and highlight the role that different features of contacts patterns and their dynamical aspects play during the course of the simulated outbreak. The results have important implications to identify the level of detail needed in the contact data in order to adequately and realistically inform modeling approaches applied to public health problems.

## Methods

### Data collection platform & Deployment

Contact network measurements are based on the SocioPatterns RFID platform *(28,30)*. Individuals are asked to wear a badge equipped with an active Radio Frequency IDentification (RFID) device ("tags"). RFID devices engage in bi-directional radio communication at multiple power levels, exchanging packets that contain a device-specific identificator. At low power level, packets can only be exchanged between tags within a 1-2 meters radius *(28, 30)*. This threshold has been set in order to detect a close-contact situation during which a communicable disease infection can be transmitted either following, for example, cough or sneeze, or directly by physical contact. Individuals were asked to wear the RFID badges on their chest, so that contacts are recorded only when participants face each other, as the body acts as a shield for the proximity-sensing RF signals. In addition to sensing for nearby devices, RFID tags send the locally collected contact information to a number of receivers installed in the environment, which relay this information over a local area network to a computer system used for monitoring and data storage. Proximity scans are performed at random times and each tag dispatches information to the receivers every few seconds. Time is then coarse-grained over 20 second intervals, over which face-to-face proximity can be assessed with a confidence in excess of 99% *(28, 30)*. This time scale also is also adequate to follow the dynamics of social interaction.

All communication is encrypted: from tag to tag, from tags to receivers, and from receivers to the data storage system. Contact data are stored in encrypted form, and all data management is completely anonymous. Other details on the data collection infrastructure can be found elsewhere *(28,30)*.

Participants attending the 2009 Annual French Conference on Nosocomial Infections (www.sfhh.org) were asked to wear RFID tags. Face-to-face interactions between 405 voluntary individuals among the 1,200 attendees were collected during two days of the conference (June 3rd and 4th, 2009). The data was collected from 9am to 9pm on the first day and from 8.30am to 4.30pm on the second day (periods defined as "day" in the following). Contacts were not recorded outside of these time periods (periods defined as "nights"). Attendees to the conference signed a written informed consent. The data have been collected anonymously. The ethics committee of Lyon university hospital gave the agreement for this study.

**Empirical contact networks**

In order to assess the role of the dynamical nature of the network of contacts in the spreading dynamics, we consider the network built on the explicit representation of the dynamical interactions among individuals (referred to as DYN), at the shortest available temporal resolution (20 seconds) against two benchmark networks that are built on progressively lower amounts of information available on the interactions. These are referred to as HET, and HOM.

Taking advantage of the full spatial and temporal resolution, DYN considers the empirical sequence of successive contact events collected during the congress. Each contact is identified by the RFID identification numbers of the two individuals it involves, and by its starting and ending times. The resulting network is a dynamical object that encodes the actual chronology and duration of contacts, therefore preserving the heterogeneity in the duration of contacts, as well as the causality constraints between events. The latter is particularly important for spreading processes, as it may prevent the propagation along certain sequences of interactions that would be otherwise allowed in an aggregated static representation of the contact patterns. For example, if a susceptible individual A interacts first with an infectious individual B and then with a susceptible C, a disease transmission can occur from B to A and then from A to C. If instead A meets first C and later B, A can get infected from B, but the propagation from B to A and then to C is not possible anymore.

The benchmark networks correspond to a coarse-graining of the data on a daily scale. The first one, HET, is produced for each conference day by connecting individuals who have been in contact during this conference day, thus aggregating all daily dynamical information in a single snapshot, and weighting each link by the total time the two individuals spent in face-to-face presence during the considered day. Therefore, HET includes information on the actual contacts between individuals (who has met whom) and on the total duration of these contacts (how long A has been in contact with B during the whole day), but disregards information about the temporal order of contacts. In the previous example, the transmission from A to C could take place in both situations representing the different sequences of the events. It is a daily aggregated network in which contacts are aggregated over a day but keeping the whole neighborhood structure between individuals. As the conference lasted two days, the aggregation procedure produces two such networks, one each day.

Finally, a homogeneous network (HOM) is constructed for each day by connecting individuals who were in face-to-face contact during the conference day, again aggregating all daily dynamical information in a single snapshot, but weighting each link with equal weight, corresponding to the average duration of contacts between two person that have met each other the same day in the HET network. The HOM construction may correspond to networks constructed by asking each participant to report with whom he or she has been in contact during the conference day and then estimating for how long this lasted on average. For each conference day, HET and HOM have exactly the same structure of interactions from a topological point of view, and they differ by the assignments of weights on the links.

**Generation of contact networks on longer timescales**

Since we simulate the spreading of a realistic infectious disease characterized by longer timescales than the data collection period, we introduce three different procedures to longitudinally extend the data-driven network, by preserving some of its features. The simplest procedure consists in repeating the two-day recordings. This repetition procedure, denoted by "REP" in the following, is performed for the dynamical sequence of contacts, and consistently for the set of daily HET and HOM networks. In this simple procedure, each attendee repeated again and again the same contacts for each simulated sequence of two days, i.e., always meets the same set of other attendees, in the same order and during the same duration. While this procedure yields a realistic contact pattern for each single day, as it uses only empirical data, such a "deterministic" repetition is rather unrealistic as time goes on. We therefore consider two additional procedures that improve this limitation.

The first one, "RAND-SH", consisted in producing two-day sequences by randomly reshuffling the participant's identities, as given by their tags IDs. The overall sequence of contacts was preserved, but each contact occurs between different attendees from one two-day sequence to the next. DYN networks were then constructed as before by taking into account the 20 seconds temporal resolution, while HET and HOM networks were obtained by aggregating the data for each day, as previously explained. More realistic contact patterns are thus obtained, in a way that avoids the unrealistic repetition of interactions between individuals. However, the RAND-SH procedure completely erases correlations between the contact patterns of an attendee in successive two-day sequences, which is also unrealistic. The analysis of the empirical contact networks shows indeed that a correlation exists between the number of contacts of an attendee in the first and second conference days and also that a fraction of contacts are repeated from one day to the next. We therefore design a third procedure for the generation of synthetic contact patterns starting from the two-day sequence ("CONSTR-SH") that constrains the reshuffling to preserve the correlations between the attendees' social activity as well as the same fraction of repeated contacts between successive days. The description of the data extension procedure 'CONSTR-SH' is shown in the Additional Material.

It is important to note that in all cases we preserve the time frame during which data was collected, since no collection occurred outside the conference premises. For this reason, each individual was considered as isolated during the "nights" periods in the DYN network. We therefore introduce such "nights" in the HET and HOM networks by "switching off" the links (i.e., considering individuals as isolated) during these periods, thus resembling the circadian pattern encoded in the empirical data.

**Epidemiological model**

We consider a simple SEIR epidemic model for the simulation of the infectious disease spread in the population under study, in which no births, deaths or introduction of individuals occurred. Individuals are each assigned to one of the following disease states: susceptible (S), exposed (E), infectious (I), recovered (R). The model is individual-based and stochastic. Susceptible individuals may contract the disease with a given rate when in contact with an infectious individual, and enter the exposed disease state where individuals are infected but not yet infectious.

Exposed individuals enter the I class at a rate $\sigma$ with $\sigma^{-1}$ representing the average latent period of the disease. Infectious individuals can transmit the disease during their infectious period whose average duration is equal to $v^{-1}$. After this period, they enter the recovered compartment, acquiring permanent immunity to the disease.

In order to compare simulation results obtained from the three different networks, we need to adequately define the rate of infection on a given infectious-susceptible pair, depending on the definition of the networks themselves. Let us define $\beta$ as the constant rate of infection from an infected individual to one of his/her susceptible contacts on the unitary time step $dt$ of the process. Given then two individuals, an infectious A and a susceptible B who are in contact during the unitary time step, the probability of B becoming infected during this period is given by $\beta dt$. In order to obtain the same average infection probability in the HET and HOM networks over an entire day (day + night), the weights on such networks need to be rescaled by $W_{AB}/\Delta T$, defined as the ratio between the total sum of the duration of all contacts between A and B in a day and the effective duration of the day (i.e. total time during which the links in the daily networks are considered active, discarding the "nights"). Therefore the infection probability between A and B during the time step $dt$ is $\beta W_{AB} \, dt/\Delta T$ for the HET network, and $\beta <W> dt/\Delta T$ for the HOM network (with $<W>$ being the average weight of the links in the HET network).

We consider two different disease scenarios for the spreading simulations on all networks under study. In particular, the duration of the average latency period, average infectious period and transmission rates assume the following values: (i) $\sigma^{-1}=1$ days, $v^{-1}=2$ days and $\beta=3.10^{-4}$ s$^{-1}$ (very short incubation and infectious periods); (ii) $\sigma^{-1}=2$ days, $v^{-1}=4$ days and $\beta=15.10^{-5}$ s$^{-1}$ (short incubation and infectious periods). These sets of parameter values are chosen in order to maintain the same value of $\beta/v$, which is the biological factor responsible for the rate of increase of cases during the epidemic outbreak, while changing the global timescales of incubation and infectious periods, and assessing the role played by the social factors embedded in the contact patterns. These values correspond to short incubation and infectious periods, in order to minimize the consequences of the arbitrariness in the construction procedures of long data sets as described previously. Each simulation started with a single randomly chosen infectious individual, with the rest of the population in the susceptible state.

**Analysis of the empirical contact networks and of the simulation results**

To describe the empirical contact networks, we reported the numbers of contact, the average duration of contacts, the average degree of a node defined as the number of distinct persons encountered by the corresponding individual, the average clustering coefficient, which describes the local cohesiveness, the average shortest path defined as the average number of links to cross to go from one node to another one, and the correlation between the properties of the nodes in the aggregated networks of the first and second conference day. For this analysis, we measure Pearson correlation coefficients between the degree of an individual in the first and second day and between the time spent in interaction in the first and second day.

The comparison of the epidemic outbreaks in the three networks under study is performed by analyzing a variety of quantities including the final size of the epidemic, the number of infectious individual during the epidemic peak, the time of the peak, and the epidemic duration.

Given the aim of assessing the role of the contact patterns, their dynamical aspect and possible different reductions of this information on the spreading phenomena, we also estimate the reproductive number $R_0$, defined as the expected number of secondary infections from an initial infected individual in a completely susceptible host population (*1*). Several methods can be used to compute this quantity (*40, 41*), possibly yielding different estimates (*42*) for the same epidemiological parameters Here we compute the value of $R_0$ as the average, over different realizations, of the number of secondary cases from the single initial randomly chosen infectious individual. Average $R_0$ values and variances are then compared for the 3 networks (DYN, HET and HOM) and 3 data extension procedures (REP, RAND-SH and CONSTR-SH) under study.

## Results

Overall, 28,540 face-to-face contacts between 405 attendees of a two-day conference were recorded. Figure 1 reports the probability distribution of the duration of these contacts. The average duration was of 49 seconds with large variations (the standard deviation is 112 seconds), meaning a large number of contacts of brief duration, few contacts of long duration, and a broad tail, suggesting that no typical contact duration can be defined. Statistical distributions of the number and duration of contacts as well as of the link weights were similar from one day to the next, although the two daily contact networks are obviously not identical.

In the daily contact networks, the average degree of a node was close to 30, with a distribution decaying exponentially for large numbers. The average clustering coefficient was 0.28, much larger than the average value of 0.07 obtained for a random network with the same size and average degree. The network was also a small-world, with an average shortest path of 2.2.

The links weights were on the other hand broadly distributed, with an average cumulated duration of the interaction between two attendees of 2 minutes. The total duration spent in contact by any attendee was also broadly distributed, with an average of 1h15mn. The Pearson correlation coefficient was 0.37 between the degree of an individual in the first and second day, and 0.52 between the total time spent in interaction in the first and second day. The fraction of repeated contacts in the second day with respect to the first was of 12%, and was independent from the degree.

Figure 2 reports the distributions of $R_0$ for the three networks, for the REP procedure. In all cases, the number of secondary cases from the initial seed of the single infectious individual ranges from 0, corresponding to the most probable event of no outbreak, to around 20-25 individuals. Figure 3 and supplementary table 1 give the average values and the variances obtained for the estimation of $R_0$ depending on the scenarios and the network type. In all scenarios, higher values of $R_0$, together with larger variances, are observed in the HOM network compared to the HET and DYN networks.

Figure 4 shows the distribution of the final number of cases for the three networks and the REP data extension procedure. A high probability of rapid extinction of the pathogens spread is observed, corresponding to a small number of individuals who become infected. This is slightly smaller in the HOM case compared to the HET and DYN networks. On the contrary, when the epidemic starts, the final number of cases is high, and it is larger in the HOM case as compared to the HET and DYN networks. Intermediate cases with limited propagation are rare.

Table 1 and supplementary figure 4 summarize the distribution of the final number of cases for the three networks for the various parameters of the SEIR model and in the various data extension scenarios. For all cases, and independently from the procedure adopted for extending the two-day data set, the probability of extinction is lower for the HOM cases compared to the HET and DYN networks. In case of propagation, the final size is higher in the HOM network compared to the HET and DYN networks. Propagation over HET and DYN networks leads to similar extinction probability and final number of cases. The final numbers of cases are also rather close in the two disease scenarios.

Figure 5 and supplementary table 2 display the peak times of the spreading in the various cases. In most cases, the epidemic peak is reached on average first for the spreading on the HOM network. The differences between the peak times are however small: even the simulations on the network with the least information give a good estimate of the peak time obtained when including the full information on the contact patterns.

Figures 6 and 7 display the temporal behavior of the spreading, through the evolution in time of the number of infectious and recovered individuals, for the different data extension procedures and for the two sets of SEIR parameters. Symbols represent the median values and lines represent the 5th and 95th percentiles of the number of infectious and recovered individuals. In all cases, the spreading on the HOM network evolves slightly faster, and reaches a significantly larger number of individuals, while spread on HET and DYN present very similar characteristics.

Figures 5 to 7 also highlight interesting differences in the results of simulations on data sets extended with different procedures: the spread is slightly slower in the RAND-SH case, but lasts longer, and as a result the final number of cases $R\infty$ is larger. In fact, we have systematically $R\infty(REP) < R\infty(CONSTR-SH) < R\infty(RAND-SH)$: the more the identities of the tags are shuffled, the more efficient the spreading.

## Discussion

Using a recently developed data collection technique deployed in a two-day conference involving 405 volunteers, we have provided measurements of the dynamics of contact (close face-to-face) interactions between individuals during such a social event. We have used the data to compare the simulated spreading of communicable diseases on this dynamic network and on two heterogeneous and homogeneous networks obtained by aggregating the dynamic network at two distinct levels of precision. To compensate for the relatively short duration of the data set (two days), we have put forward different procedures to provide contact networks for an extended time period during which the spread of an infectious disease can be simulated.

The broad distributions of the various characteristics of the network analyzed in this study have been, as expected, also observed in other contexts (*30, 36, 37*). In the present case, we obtained results which are statistically similar to other conference interaction networks, as reported in (*30, 36*). As also described in (*30, 36*), the resulting picture is not characterized by the presence of "superspreaders" if defined in terms of the number of different persons contacted, although it was less clear if the cumulated interaction time was taken into account.

In the three networks, disease extinction occurs as frequently (between 36% and 47%) as large outbreaks (between 34% and 49%) and outbreaks are rather explosive (attack rate between 51% and 80%) which is consistent with previous work (*4*). A strong difference in the spreading process is observed between the HOM network that does not include any information on the heterogeneity of contact durations nor on the dynamical aspect and the two other networks, with a systematically larger number of infected individuals for the HOM network. This result implies that heterogeneity in the contact durations between individuals is associated with a lower spread of transmission. That suggests that one individual who does not spend her/his time equally between her/his contacts effectively reduces the spreading routes (*12,15*). Disregarding the heterogeneity of contact durations can lead to strong differences in the estimation of the number of cases, suggesting that information on the daily cumulated contact time between individuals gives crucial information for correctly modeling disease spreading. Interestingly however, the peak time is only slightly changed in the HOM network, showing that even rather limited information can yield good estimates of the epidemic timescales.

The comparison between the spreading on the HET and DYN networks provides us with insights on whether temporal constraints due to the precise sequence of the contacts may impact the propagation of diseases. Given two individuals, the overall expected probability of a transmission occurring during the interval $\Delta T$ is indeed the same in both cases (i.e., $\beta W_{AB}$) so the only difference is that the contact is not continuously present in the DYN network, but it may be intermittent and repeated only during the actual recorded contacts. This introduces time constraints on the paths that the infectious agent can follow between individuals in the DYN network, which may slow down the spreading on the DYN network, compared to the HET network.

This slowing down and the differences in the final number of cases between the HET and DYN are however too small to be relevant for the simulations investigated here. The similarity between the spreading behaviors on HET and DYN was observed independently from the different procedures used to extend the initial two-day data set. These procedures create successive artificial "days" which differ from each other by various amounts, i.e., with a different amount of repetition of contacts from one day to the next. The robustness of the comparison between HET and DYN indicates therefore that it is due to the strong difference between the timescales considered for propagation, which are of the order of days, and the temporal resolution and the contact durations, respectively of 20 seconds and of the order of minutes, up to a few hours. The information contained in the total time spent in contact by each pair of individuals is in this context sufficient to describe precisely the propagation pattern, as described by the peak time and the final number of cases. Therefore, for the simulation of diseases such as those in this study, contact information at a daily resolution may be enough to characterize the spreading, and the precise order of the sequence of contacts could not be needed. This would however not be the case for extremely fast spreading processes, as shown in (*36*); this implies that there is a crossover between the two regimes, which will be the subject of future investigations.

The difference between the results obtained for the different procedures REP, RAND-SH and CONSTR-SH finally shows the importance of the knowledge of the respective fractions of repeated and new contacts between successive days (*8, 12, 43*). Repeated encounters favor propagation, so that the REP procedure leads to faster spread at short times, but contacts between different individuals from one day to the next favor propagation across the network, so that the RAND-SH procedure leads in the end to a larger attack rate.

Compared to other approaches (*11, 26, 27*), the data collection method used in this study makes it possible to gather information on actual face-to-face contacts, with high temporal and spatial resolutions (*28,30,36*). This gives access to the precise durations as well as the time and order of the successive contacts between individuals, fully representing the corresponding heterogeneity and the causality constraints in the chain of transmission.

Unsupervised data collection systems based on RFID infrastructures, such as the one presented here (*28, 30, 37*), present some caveats that need to be discussed. First, individuals are not followed outside of the zone covered by RFID readers, so that contacts between participants that occur during the day outside of the area covered by the RFID readers are not monitored. This leads to an underestimation of the number of contacts, and therefore of the spreading possibilities. Moreover, periods of "nights" represent a proportion of 56% of time during which individuals are assumed to be isolated. This may artificially increase the probability of extinction if the contagiousness period of an infected individual ends during these periods, precluding further transmission. This issue may be solved by upcoming technological improvements that will allow operating the RFID sensing layer in a fully distributed fashion with on-board storage on the devices, i.e., such RFID tags will register and store contacts even if they are not close to RFID readers.

Another issue, well known in the field of social networks, is due to the partial sampling of the population. Among the 1,200 attendees to the conference, 405 (34%) have participated in the data collection. Only these attendees are taken into account in the spreading model, while they were in fact also in contact with the remaining attendees. Previous investigation (*30*) has shown that for a broad variety of real-world deployments of the RFID proximity-sensing platform as used in this study, the behavior of the statistical distributions of quantities such as contact durations is not altered by unbiased sampling of individuals. However, spreading paths between sampled attendees involving unsampled attendees may have existed, but are not taken into account. This effect may lead to an underestimation of the spreading, and future work will focus on a quantification of such possible biases, for instance through bootstrapping procedures. In addition, it is possible that volunteering participants may introduce a systematic bias in the sampled population concerning their interaction behavior, as they self-select to participate to the experiment. The assessment of this effect would however require independent data sources for monitoring unsampled individuals, inevitably limiting the size of populations and settings because of logistics constraints. Though interesting for the understanding of social behavior, such a study would need to be specifically designed and tailored to the research question, thus going beyond the aim of the present study. Another interesting perspective would be to compare and integrate the results of unsupervised contact measurements with the results of simultaneously performed surveys- or diaries-based inquiries.

Finally, the limited period of time (two days) of data collection made it necessary to generate artificially longer data sets by different procedures, in order to model the spreading of pathogens on realistic timescales. Deployment of the measuring infrastructure on much longer timescales is planned, in order to validate such generation procedures and to measure their effect.

## Conclusions

In spite of these limitations, the present study emphasizes the effects of contacts heterogeneities on the dynamics of communicable diseases. On the one hand, the small differences between simulated spreading on HET and DYN networks shows that taking into account the very detailed actual time ordering of the contacts between individuals, at a time resolution of minutes, does not seem essential to describe the spreading on the timescale of several days or weeks. On the other hand, the strong differences observed with the spreading on the HOM network underline the need to include detailed information about the contact duration heterogeneity (compared to an assumption of homogeneity) to model disease spread, as also found in (*12, 13*) for simulations of spreading dynamics based on diary-based survey data. Results for the different procedures for the extension of data show also how the rate of new contacts is a very important parameter (*8, 12, 43*). Overall, the combined comparison of the spreading processes simulated on the HET, DYN and HOM networks, and using the different data extension procedures, give an important assessment of the level of details concerning the contact patterns of individuals needed to inform modeling frameworks of epidemics spreading.

In this context, a data collection infrastructure such as the one used in this study appears to be very effective, as it gives access to the level of information needed, and also allows the simulation of very fast spreading processes characterized by timescales comparable to the ones intrinsic to the social dynamics, where even the precise ordering of contacts events becomes crucial. These measurements should be also extended to other contexts were individuals are closely interacting in different ways, such as workplaces, school or hospitals (*44*). More experimental works is needed to collect data over longer time periods, in particular to understand better how data sets limited in time can be artificially extended to yield realistic data sets, on various samples of individuals and in various locations. The results of these approaches could be helpful to anticipate the impact of preventive measures and contribute to decisions about the best strategies to control the spreading of known or emerging infections.

## Competing interests

The authors declared no competing interests.

## Authors' contributions

JS, NV, AB, CC, VC, LI, CR, JFP, WVdB and PV conceived and designed the experiments. NV, AB, CC, CR, JFP, NK, WVdB and PV performed the data collection. JS, NV, AB, CC, VC, LI and JFP analyzed the data. JS, NV, AB, CC, VC, LI, JFP and PV wrote the paper. All authors read and approved the final manuscript.

## Acknowledgements

We acknowledge the contribution of all partners of the SocioPatterns project. We are grateful to the organizers of the conference of "Société Française d'Hygiène Hospitalière (SFHH)" .VC is partially supported by the ERC Ideas contract ERC-2007-Stg204863 (EPIFOR) and by the FET projects Epiwork and Dynanets. LI is partially supported by the FET project Dynanets. This project was partly supported by "La Société Française d'Hygiène Hospitalière" and GOJO France. This study was partly supported by a grant of the "Programme de Recherche, A(H1N1)" co-ordinated by the "Institut de Microbiologie et Maladies Infectieuses". We thank all the attendees of the conferences who volunteered to participate in the data collection.

# Figures

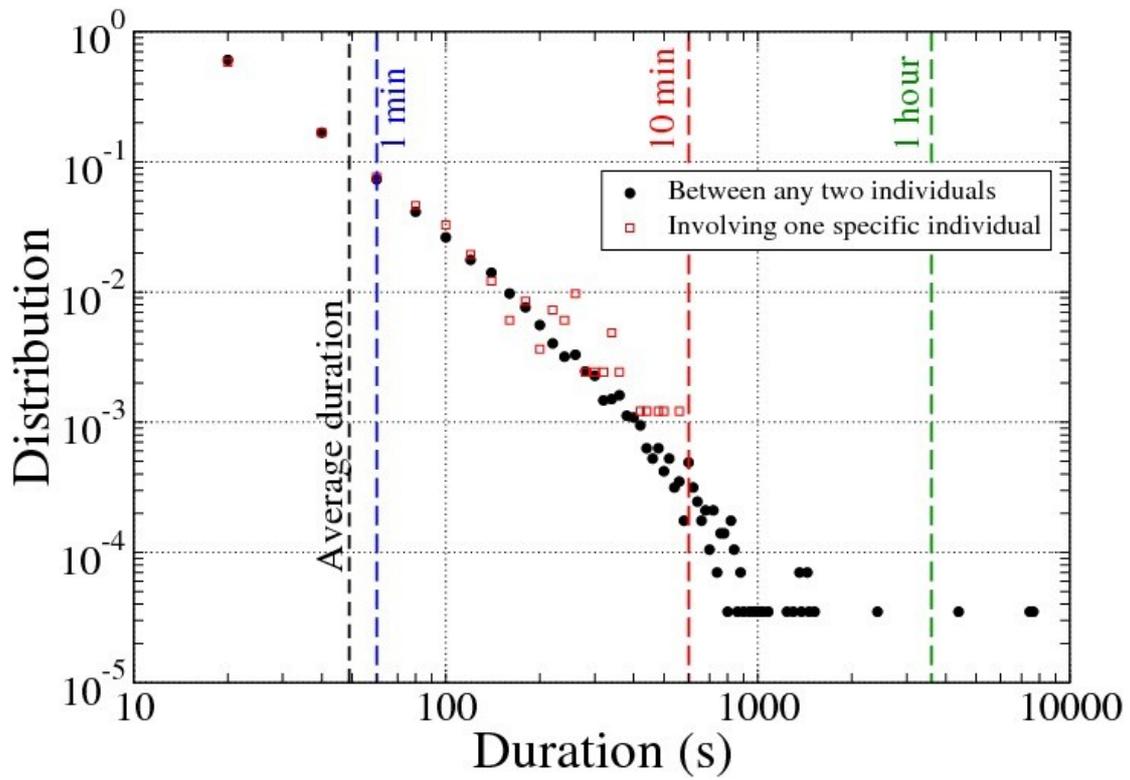

Figure 1 – Distribution of the contact duration between any two individuals on a log-log scale. The average duration is 49 seconds, with a standard deviation of 112 seconds.

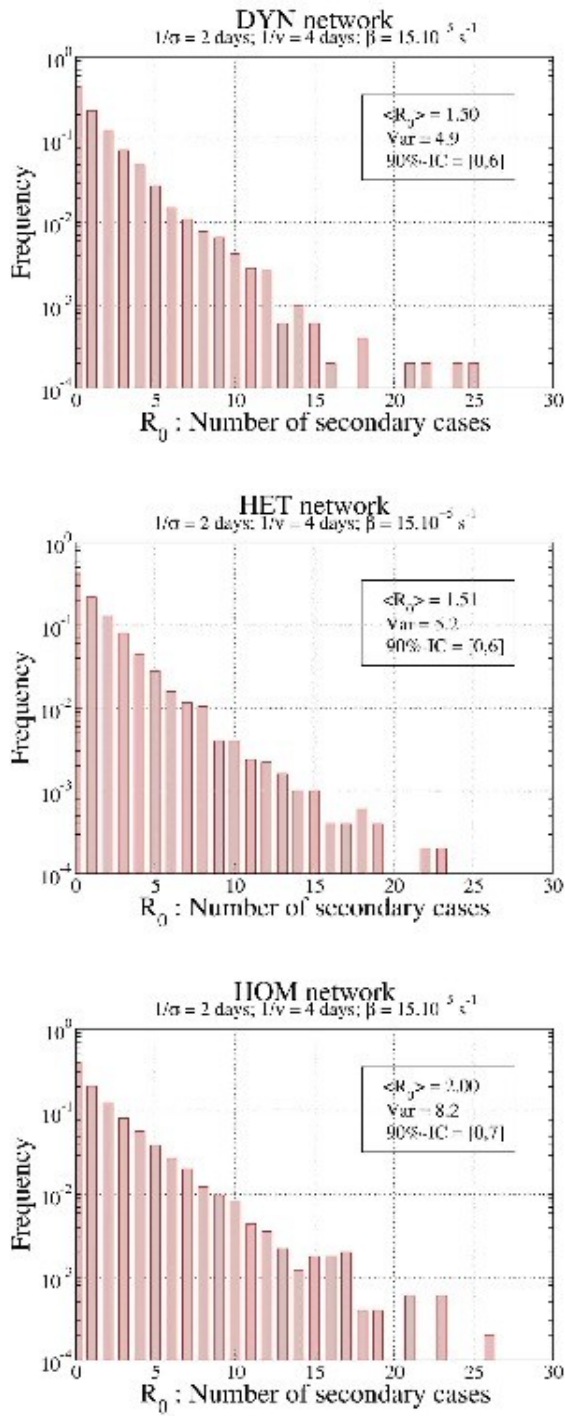

Figure 2 – Distribution of $R_0$ for the HOM, HET and DYN networks with the parameters $s^{-1}$ =2 days, $n^{-1}$ = 4 days and $\beta$ = 15.10$^{-5}$ s$^{-1}$ (short latency, short infectiousness), in the REP procedure.

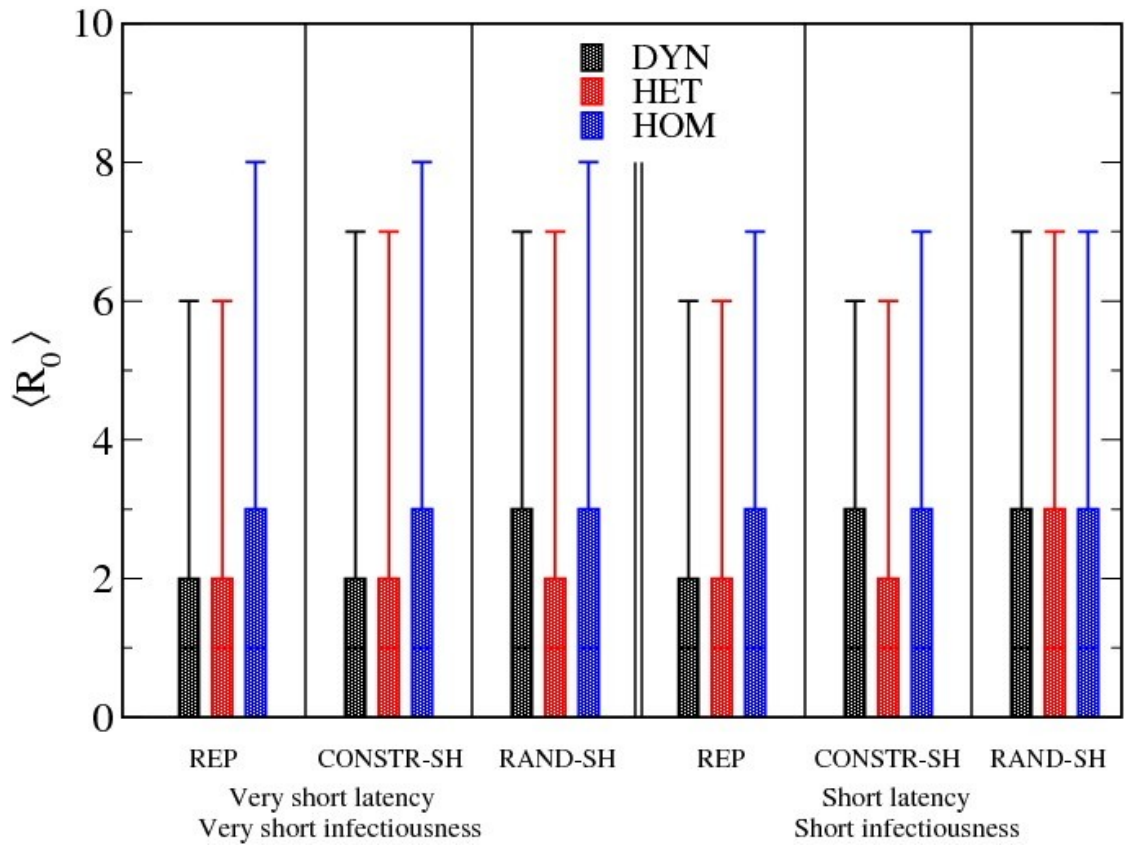

Figure 3 – Boxplots showing the distributions of $R_0$ according to the different scenarios and network types. The bottom and top of the rectangular boxes correspond to the 25th and 75th quantile of the distribution, the horizontal lines to the median, and the ends of the whiskers give the 5th and 95th percentiles. Very short latency, very short infectiousness scenario: $\sigma^{-1}$=1 days, $\nu^{-1}$=2 days and $\beta$=3.10$^{-4}$ s$^{-1}$. Short latency, short infectiousness scenario: $\sigma^{-1}$=2 days, $\nu^{-1}$=4 days and $\beta$=15.10$^{-5}$ s$^{-1}$.

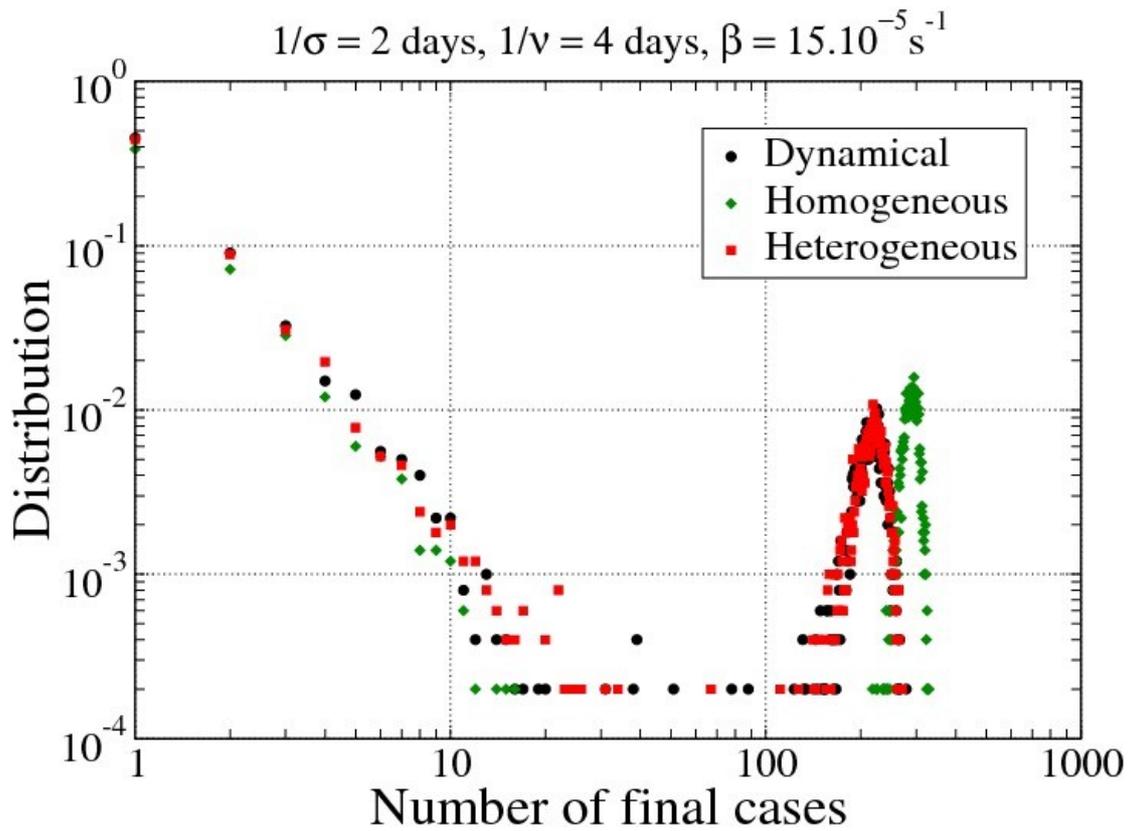

Figure 4 – Distribution of the final number of cases for the three networks with the parameters $s^{-1}$ =2 days, $n^{-1}$ = 4 days and $\beta$ = 15.10$^{-5}$ s$^{-1}$ (short latency, short infectiousness), in the REP procedure.

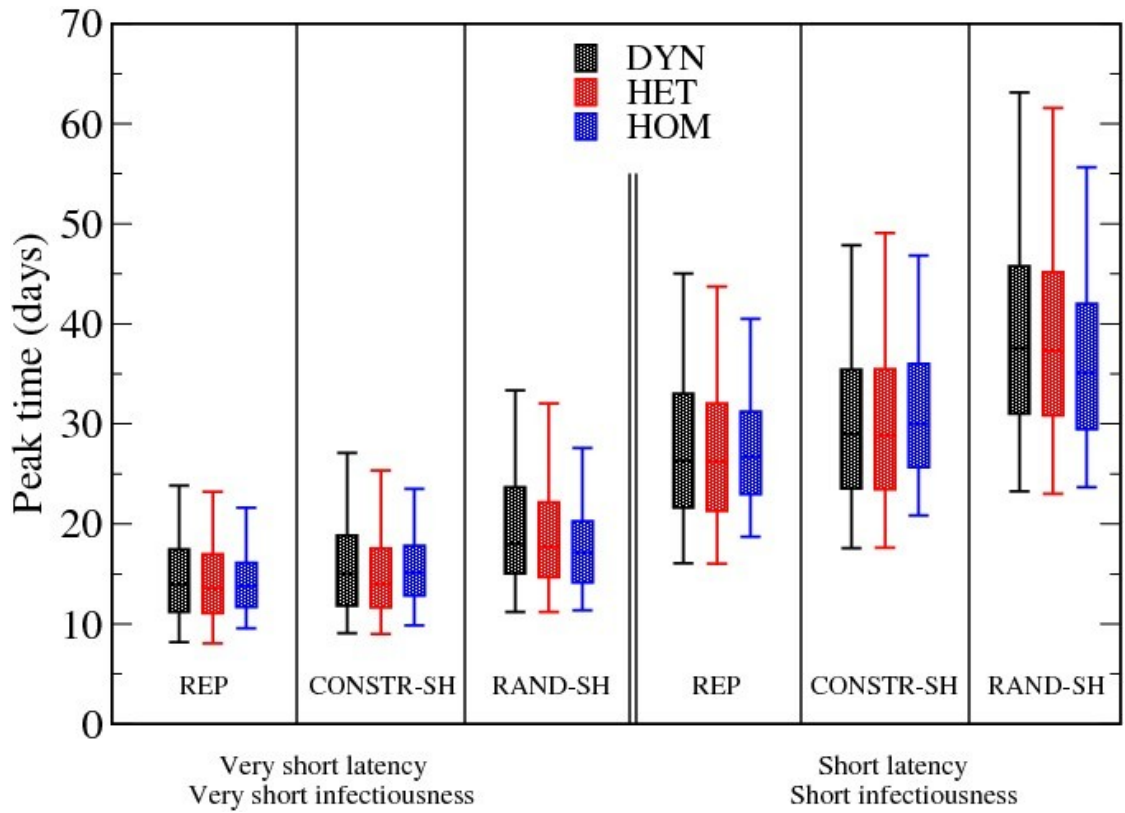

Figure 5 – Boxplots (symbols as in Fig 3.) showing the distributions of the prevalence peak time $t_{peak}$ according to the different scenarios and network types. Only runs with AR>10% are taken into account. Very short latency, very short infectiousness scenario: $\sigma^{-1}$=1 days, $\nu^{-1}$=2 days and $\beta$=3.10$^{-4}$ s$^{-1}$. Short latency, short infectiousness scenario: $\sigma^{-1}$=2 days, $\nu^{-1}$=4 days and $\beta$=15.10$^{-5}$ s$^{-1}$.

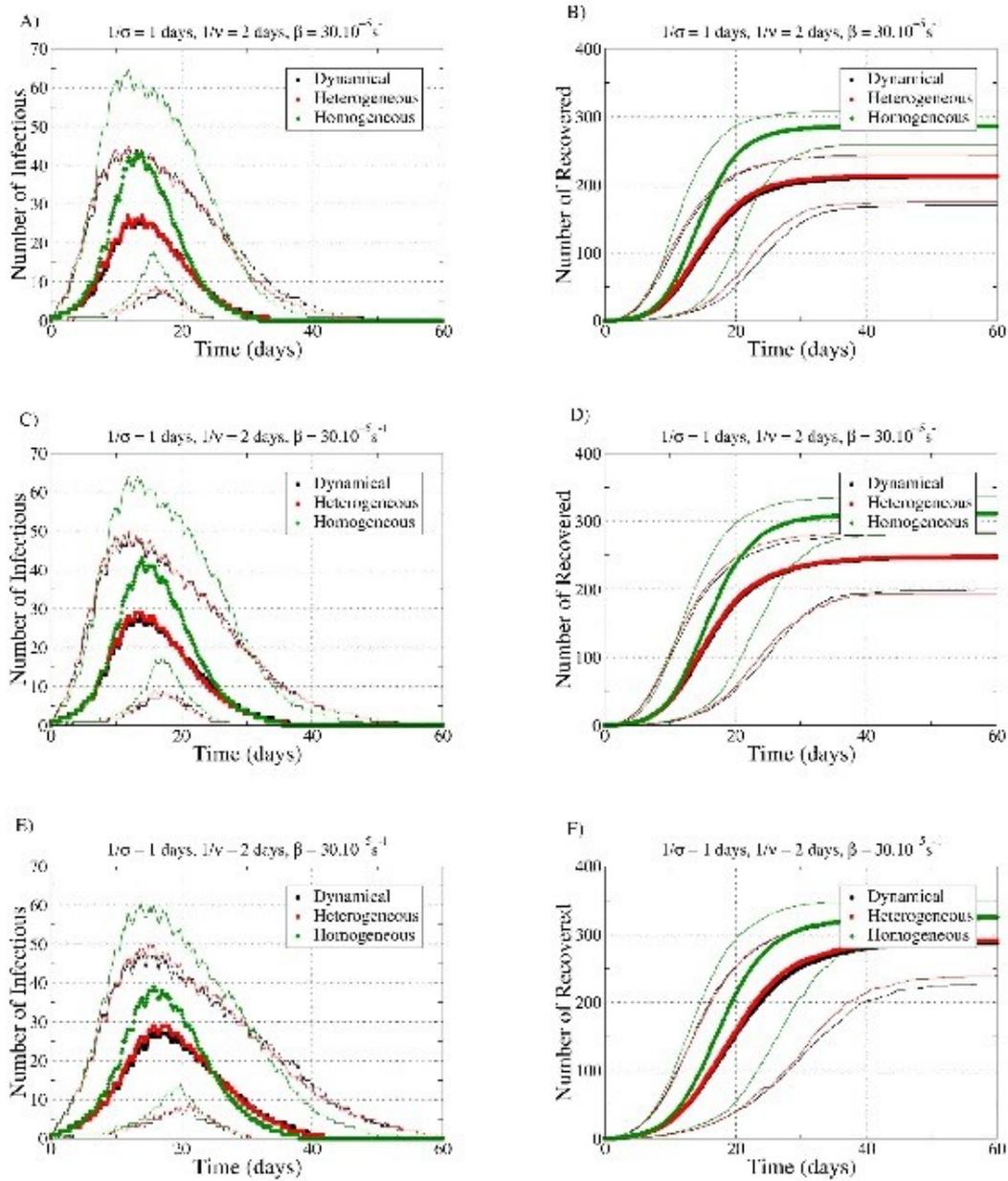

Figure 6 – Temporal evolution of the spreading process for the three networks with the parameters $\sigma^{-1}$ =1 days, $v^{-1}$ = 2 days and $\beta$ = 30.10$^{-5}$ s$^{-1}$ (very short latency, very short infectiousness). Panels A, C, E give the evolution of the number of infectious individuals, while panels B, D, F show the number of recovered. Panels A, B correspond to the REP

procedure, panels C, D to the CONSTR-SH procedure, and panels E, F to the RAND-SH one. Only runs with AR>10% are taken into account. Symbols represent the median values and lines represent the 5th and 95th percentiles of the number of infectious and recovered individuals.

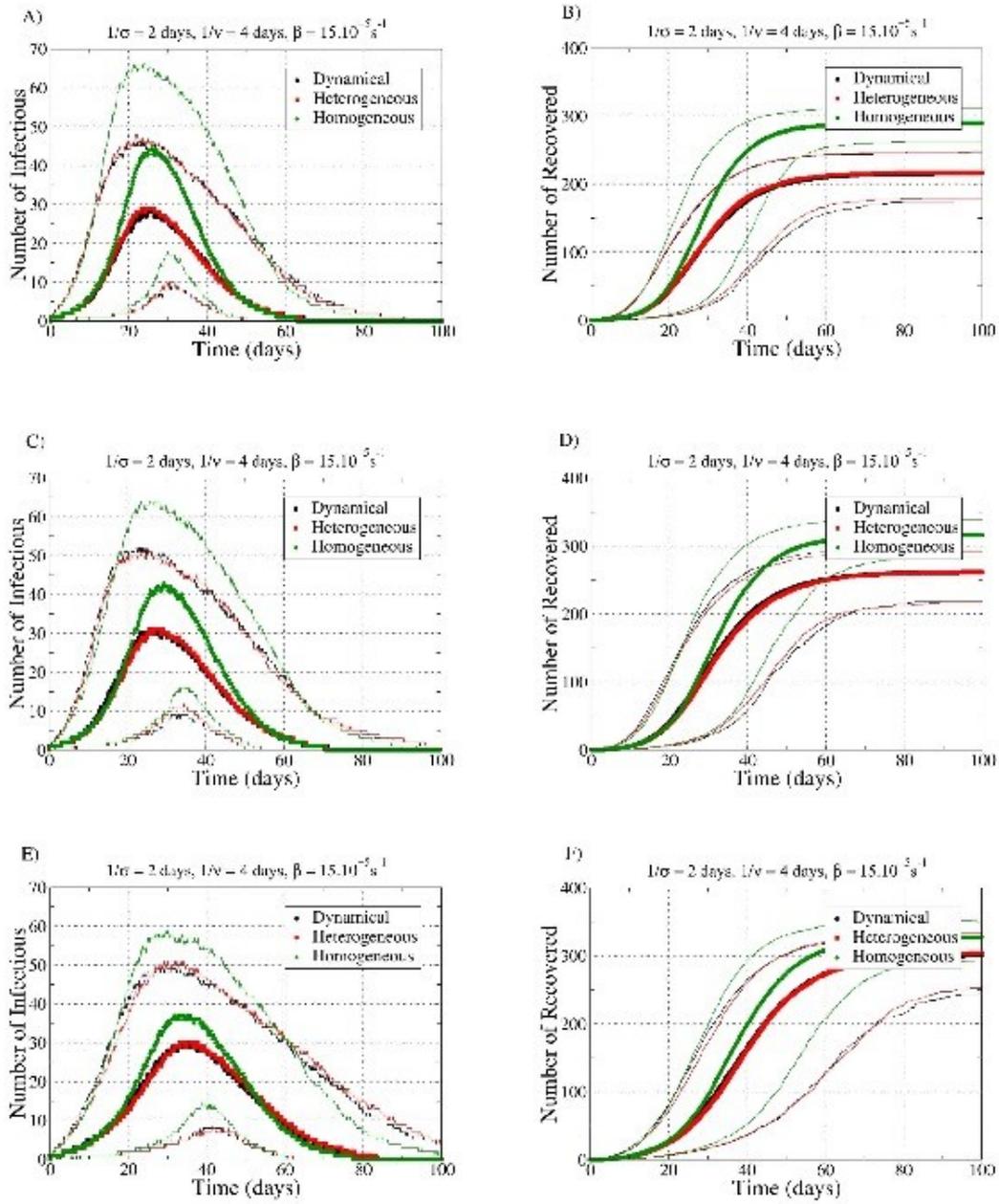

Figure 7 – Same as Fig. 4, with $\sigma^{-1}$ =2 days, $v^{-1}$ = 4 days and β = 15.10$^{-5}$ s$^{-1}$ (short latency, short infectiousness).

# Tables

**Table 1** – Distribution of the final number of cases for the three network types according to the four scenarios (5000 runs, dynamic contact network of 405 participating attendees). 90%-CI, 90% confidence interval.

| Scenarios | Parameters | Network | # of runs | % of run with no secondary cases | 1 to 10 final cases (AR*≤2.5%) | | | 11 to 40 final cases (2.5%<AR≤10%) | | | More than 40 final cases (AR>10%) | | |
|---|---|---|---|---|---|---|---|---|---|---|---|---|---|
| | | | | | % run | Average # cases | 90%-CI | % run | Average # cases | 90%-CI | % run | Average # cases | 90%-CI |
| Very short latency Very short infectiousness REP | $1/\sigma = 1$ days $1/\nu = 2$ days $\beta = 3.10^{-4}$ s$^{-1}$ | DYN | 5000 | 47.3 | 18.2 | 2.3 | [1,6] | 0.7 | 15.9 | [11,22] | 33.8 | 208 | [169,242] |
| | | HET | 5000 | 46.4 | 17.7 | 2.4 | [1,7] | 0.8 | 17.9 | [11,32] | 35.2 | 210 | [171,243] |
| | | HOM | 5000 | 41.7 | 11.7 | 2.2 | [1,6] | 0.2 | 16.6 | [11,30] | 46.3 | 285 | [257,310] |
| Short latency Short infectiousness REP | $1/\sigma = 2$ days $1/\nu = 4$ days $\beta = 15.10^{-5}$ s$^{-1}$ | DYN | 5000 | 45.3 | 17.0 | 2.2 | [1,7] | 0.4 | 18.3 | [11,38] | 37.3 | 214 | [178,246] |
| | | HET | 5000 | 44.4 | 16.4 | 2.2 | [1,6] | 0.6 | 16.8 | [11,27] | 38.6 | 216 | [178,248] |
| | | HOM | 5000 | 38.7 | 13;2 | 2.1 | [1,6] | 0.1 | 13.2 | [11,15] | 48.1 | 288 | [262,310] |
| Very short latency Very short infectiousness RAND-SH | $1/\sigma = 1$ days $1/\nu = 2$ days $\beta = 3.10^{-4}$ s$^{-1}$ | DYN | 5000 | 44.8 | 19.4 | 2.8 | [1,8] | 2.2 | 17.9 | [11,31] | 33.6 | 278 | [223,319] |
| | | HET | 5000 | 45.4 | 18.5 | 2.6 | [1,7] | 1.6 | 17.6 | [11,30] | 34.5 | 284 | [241,322] |
| | | HOM | 5000 | 39.9 | 14.3 | 2.6 | [1,7] | 0.8 | 15.7 | [11,28] | 45.0 | 324 | [291,350] |
| Short latency | $1/\sigma = 2$ days | DYN | 5000 | 40.6 | 18.6 | 2.7 | [1,8] | 1.4 | 19.2 | [11,31] | 39.4 | 297 | [254,331] |
| | | HET | 5000 | 39.5 | 18.0 | 2.7 | [1,8] | 1.3 | 16.7 | [11,30] | 41.2 | 300 | [259,333] |

| Scenarios | Parameters | Network | # of runs | % of run with no secondary cases | 1 to 10 final cases (AR*≤2.5%) | | | 11 to 40 final cases (2.5%<AR≤10%) | | | More than 40 final cases (AR>10%) | | |
|---|---|---|---|---|---|---|---|---|---|---|---|---|---|
| | | | | | % run | Average # cases | 90%-CI | % run | Average # cases | 90%-CI | % run | Average # cases | 90%-CI |
| Short infectiousness Very short latency RAND-SH | $1/\nu = 4$ days $\beta = 15.10^{-5}$ s$^{-1}$ | HOM | 5000 | 35.9 | 15.7 | 2.5 | [1,7] | 0.9 | 17.0 | [11,31] | 47.5 | 325 | [293,352] |
| | | DYN | 5000 | 45.4 | 17.7 | 2.4 | [1,7] | 1.0 | 17.0 | [11,28] | 35.8 | 240 | [194,278] |
| Very short infectiousness CONSTR-SH | $1/\sigma = 1$ days $1/\nu = 2$ days $\beta = 3.10^{-4}$ s$^{-1}$ | HET | 5000 | 46.8 | 16.5 | 2.4 | [1,7] | 0.8 | 19.0 | [11,33] | 35.9 | 245 | [202,282] |
| | | HOM | 5000 | 39.8 | 13.3 | 2.3 | [1,6] | 0.7 | 15.4 | [11,21] | 46.2 | 308 | [278,334] |
| Short latency Short infectiousness CONSTR-SH | $1/\sigma = 2$ days $1/\nu = 4$ days $\beta = 15.10^{-5}$ s$^{-1}$ | DYN | 5000 | 40.9 | 18.2 | 2.3 | [1,6] | 0.8 | 16.8 | [11,34] | 40.2 | 258 | [215,292] |
| | | HET | 5000 | 41.3 | 16.8 | 2.3 | [1,7] | 0.5 | 14.0 | [11,25] | 41.4 | 257 | [213,292] |
| | | HOM | 5000 | 35.7 | 14.8 | 2.4 | [1,7] | 0.4 | 15.2 | [11,21] | 49.2 | 314 | [284,339] |

**Additional Material**

**Description of the data extension procedure 'CONSTR-SH'.**

The data describes a list of contact events between pairs of individuals. Upon reshuffling of two tag identities, for instance of tags $i$ and $j$, an artificial data set is generated such that each time the tag $i$ was in contact with another tag, say with $k$, from time $t_0$ to time $t_1$, in the real data, the contact was replaced by a contact between $j$ and $k$ between times $t_0$ and $t_1$.

As explained in the main text, the empirical data set allows constructing daily aggregated contact networks. Let us denote by $f_{emp}$ the observed average fraction of repeated contacts from one day to the next: for each individual $i$, one considers the set $V(i,1)=\{j_1,j_2,...\}$ of individuals with whom $i$ has had a contact on day 1, and $V(i,2)=\{k_1,k_2,...\}$ with whom he or she has had a contact on day 1. The fraction $f_{emp}$ is then the average over all individuals of the ratio between the size of the intersection of $V(i,1)$ and $V(i,2)$, and the size of $V(i,1)$. If $f_{emp} = 0$, it means that $i$ has encountered only new individuals during the second day and if $f_{emp} = 1$, it means that $i$ has encountered exactly the same set of participants in both days.

For each reshuffling of the tags, we can aggregate the reshuffled contact data on a daily scale and create the reshuffled daily contact networks. We then computed the average fraction $f$ of repeated contacts between the empirical and the reshuffled daily aggregated networks. By constraining $f$ to be close to $f_{emp}$, we constructed reshuffled contact sequences that conserve a realistic amount of correlations between the sets of individuals encountered from one day to the next in the artificial data set.

We proceeded by the following steps:
1. Choose two tag Ids at random
2. Exchange their identities, as described above
3. Compute $f$ and $(f - f_{emp})^2$
4. Accept the exchange with a probability decreasing with $b\,(f - f_{emp})^2$, where $b$ is a parameter
5. Go back to step 1.

By tuning and increasing slowly the parameter $b$, it is then possible to produce reshufflings which have very low values of $(f - f_{emp})^2$, and thus reproduce the empirical correlations between the successive daily networks.

**Supplementary figure 1 –** Snapshot of the contact graph between the 405 attendees for the first conference day. Each node represents an attendee, and a link between two nodes corresponds to the fact that at least one contact event has been registered between the corresponding attendees.

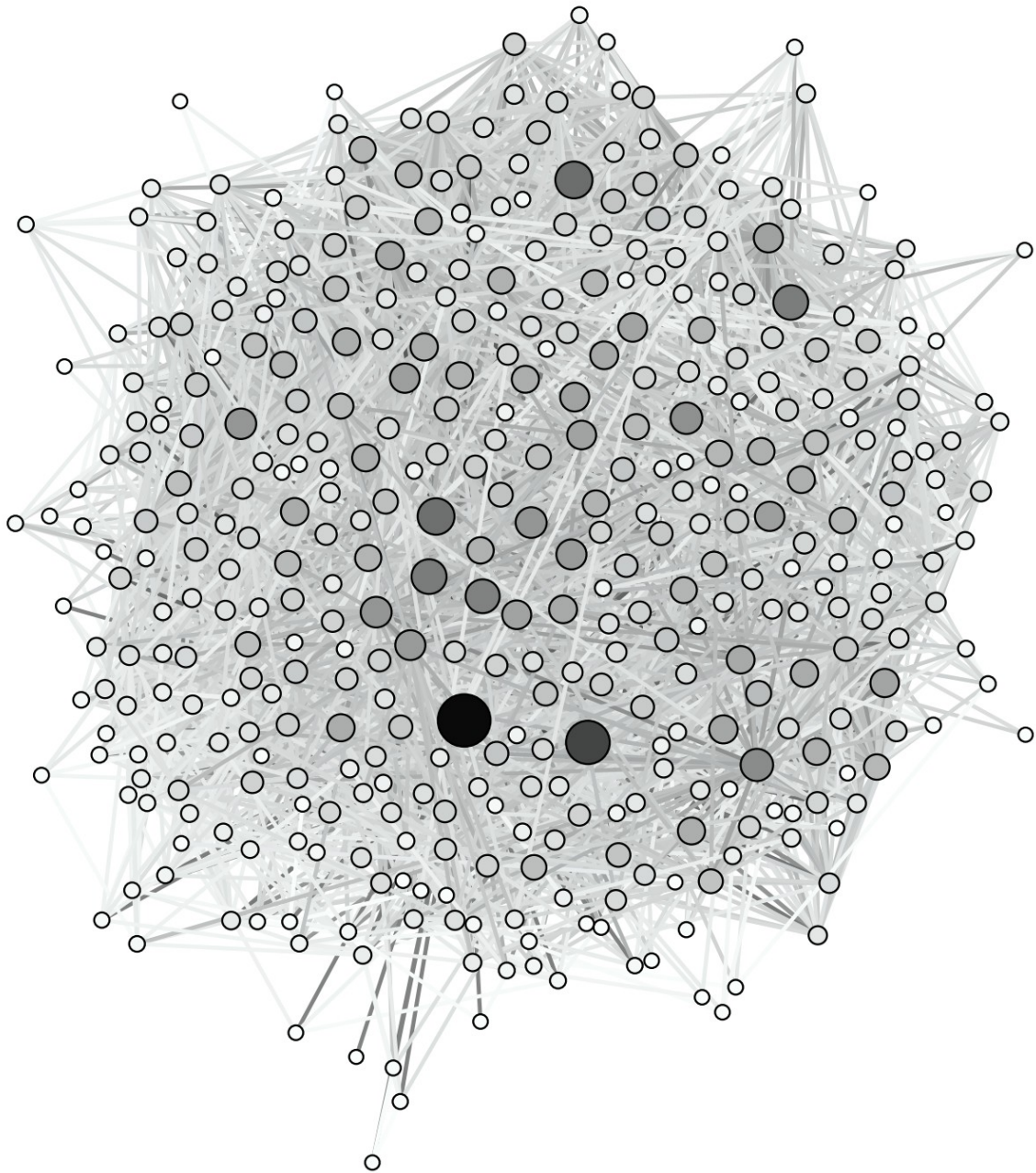

**Supplementary figure 2 –** Same as supplementary figure 1, in which only links corresponding to a cumulated time spent in proximity of at least 1mn have been kept.

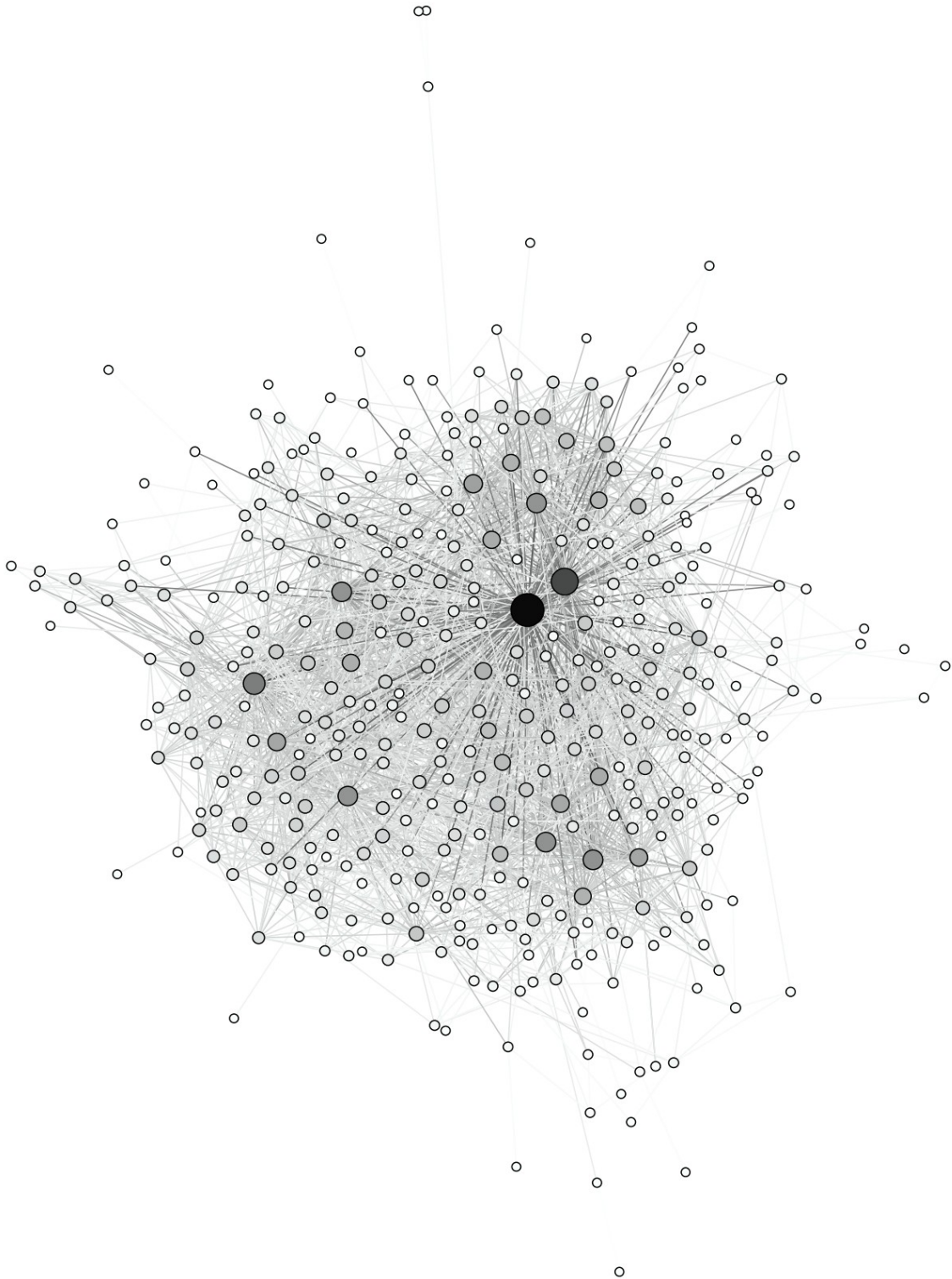

**Supplementary figure 3 –** Same as supplementary figure 1, in which only links corresponding to a cumulated time spent in proximity of at least 2mn have been kept.

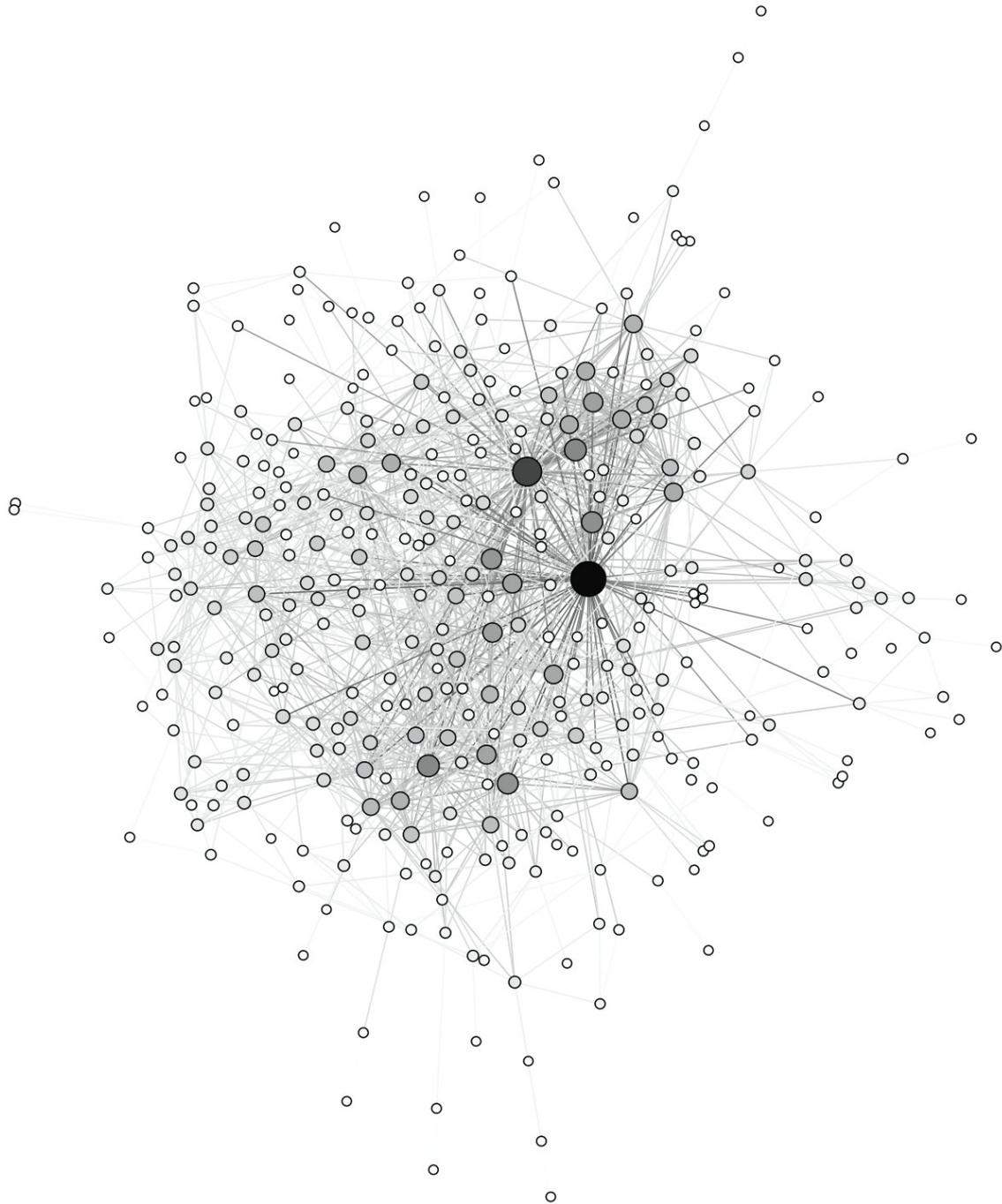

**Supplementary table 1 –** Average values, variances and 90% confidence interval (90% CI) of $R_0$ according to the different scenarios and network types.

| Scenarios | Parameters | Network | Number of runs | <$R_0$> | Variance | 90% CI |
|---|---|---|---|---|---|---|
| Very short latency Very short infectiousness REP | $1/\sigma$ = 1 days $1/\nu$ = 2 days $\beta = 3.10^{-4}$ s$^{-1}$ | DYN | 5000 | **1.55** | **6.0** | **[0,6]** |
| | | HET | 5000 | **1.46** | **5.0** | **[0,6]** |
| | | HOM | 5000 | **1.96** | **7.8** | **[0,8]** |
| Short latency Short infectiousness REP | $1/\sigma$ = 2 days $1/\nu$ = 4 days $\beta = 15.10^{-5}$ s$^{-1}$ | DYN | 5000 | **1.50** | **5.0** | **[0,6]** |
| | | HET | 5000 | **1.47** | **4.7** | **[0,5]** |
| | | HOM | 5000 | **1.93** | **7.7** | **[0,7]** |
| Very short latency Very short infectiousness RAND-SH | $1/\sigma$ = 1 days $1/\nu$ = 2 days $\beta = 3.10^{-4}$ s$^{-1}$ | DYN | 5000 | **1.99** | **6.9** | **[0,7]** |
| | | HET | 5000 | **1.70** | **5.9** | **[0,7]** |
| | | HOM | 5000 | **2.09** | **7.6** | **[0,8]** |
| Short latency Short infectiousness RAND-SH | $1/\sigma$ = 2 days $1/\nu$ = 4 days $\beta = 15.10^{-5}$ s$^{-1}$ | DYN | 5000 | **1.94** | **6.0** | **[0,7]** |
| | | HET | 5000 | **1.82** | **6.0** | **[0,7]** |
| | | HOM | 5000 | **2.03** | **6.5** | **[0,7]** |
| Very short latency Very short infectiousness CONSTR-SH | $1/\sigma$ = 1 days $1/\nu$ = 2 days $\beta = 3.10^{-4}$ s$^{-1}$ | DYN | 5000 | **1.78** | **7.1** | **[0,7]** |
| | | HET | 5000 | **1.71** | **6.5** | **[0,7]** |
| | | HOM | 5000 | **2.09** | **8.2** | **[0,8]** |
| Short latency Short infectiousness CONSTR-SH | $1/\sigma$ = 2 days $1/\nu$ = 4 days $\beta = 15.10^{-5}$ s$^{-1}$ | DYN | 5000 | **1.79** | **6.3** | **[0,6]** |
| | | HET | 5000 | **1.72** | **6.3** | **[0,6]** |
| | | HOM | 5000 | **1.98** | **6.7** | **[0,7]** |

**Supplementary figure 4** – Boxplots showing the distributions of the number of final cases when the final attack rate is larger than 10%, according to the different scenarios and network types. The bottom and top of the rectangular boxes correspond to the 25th and 75th quantile of the distribution, the horizontal line to the median, and the ends of the whiskers give the 5th and 95^th. Very short latency, very short infectiousness scenario: $\sigma^{-1}$=1 days, $v^{-1}$=2 days and $\beta$=3.10$^{-4}$ s$^{-1}$. Short latency, short infectiousness scenario: $\sigma^{-1}$=2 days, $v^{-1}$=4 days and $\beta$=15.10$^{-5}$ s$^{-1}$.

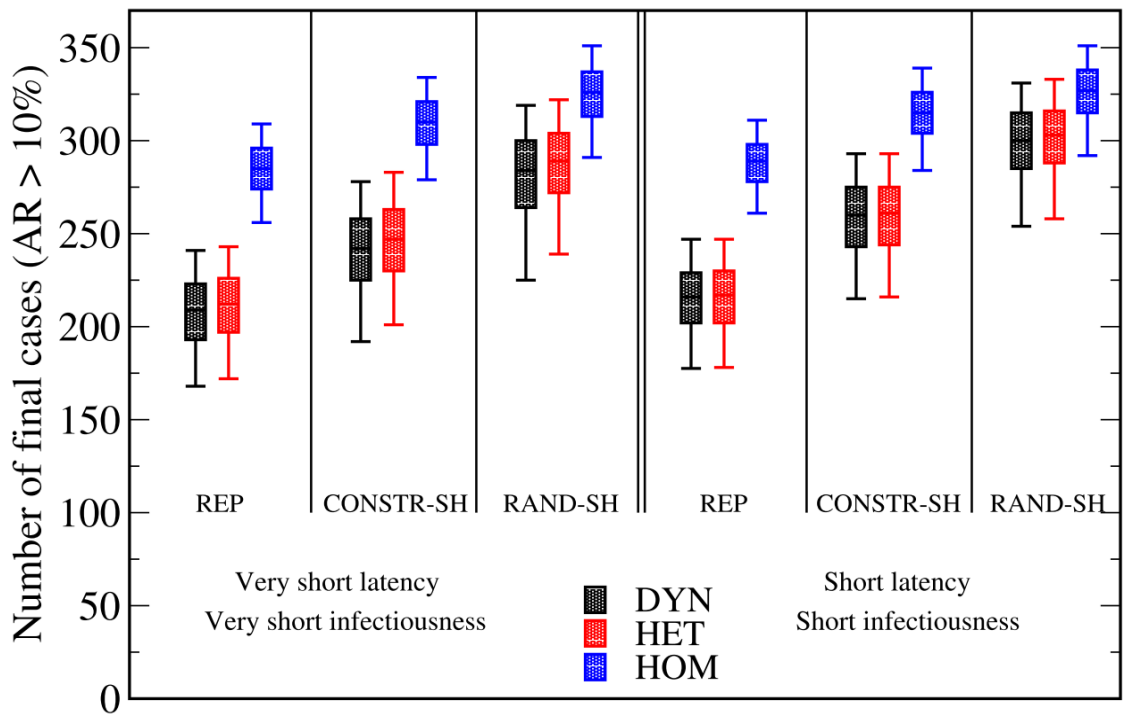

**Supplementary table 2** – Average values, variances and 90% confidence interval (90% CI) of the prevalence peak time $t_{peak}$ according to the different scenarios and network types. Only runs with AR>10% are taken into account.

| Scenarios | Parameters | Network | Number of runs | $<t_{peak}>$ days | Variance | 90% CI |
|---|---|---|---|---|---|---|
| Very short latency Very short infectiousness REP | $1/\sigma = 1$ days $1/\nu = 2$ days $\beta = 3.10^{-4}$ s$^{-1}$ | DYN | 2000 | 14.7 | 25 | [8,23] |
| | | HET | 2000 | 14.3 | 24 | [8,23] |
| | | HOM | 2000 | 14.3 | 14 | [9,21] |
| Short latency Short infectiousness REP | $1/\sigma = 2$ days $1/\nu = 4$ days $\beta = 15.10^{-5}$ s$^{-1}$ | DYN | 2000 | 28.1 | 84 | [16,45] |
| | | HET | 2000 | 27.6 | 78 | [16,43] |
| | | HOM | 2000 | 27.7 | 47 | [18,41] |
| Very short latency Very short infectiousness RAND-SH | $1/\sigma = 1$ days $1/\nu = 2$ days $\beta = 3.10^{-4}$ s$^{-1}$ | DYN | 2000 | 19.9 | 53 | [11,33] |
| | | HET | 2000 | 19.2 | 46 | [11,32] |
| | | HOM | 2000 | 17.8 | 25 | [11,27] |
| Short latency Short infectiousness RAND-SH | $1/\sigma = 2$ days $1/\nu = 4$ days $\beta = 15.10^{-5}$ s$^{-1}$ | DYN | 2000 | 39.6 | 168 | [23,63] |
| | | HET | 2000 | 39. | 148 | [23,61] |
| | | HOM | 2000 | 36.7 | 108 | [23,56] |
| Very short latency Very short infectiousness CONSTR-SH | $1/\sigma = 1$ days $1/\nu = 2$ days $\beta = 3.10^{-4}$ s$^{-1}$ | DYN | 2000 | 15.9 | 31 | [9,27] |
| | | HET | 2000 | 15.1 | 27 | [9,25] |
| | | HOM | 2000 | 15.6 | 18 | [10,24] |
| Short latency Short infectiousness CONSTR-SH | $1/\sigma = 2$ days $1/\nu = 4$ days $\beta = 15.10^{-5}$ s$^{-1}$ | DYN | 2000 | 30.4 | 97 | [17,47] |
| | | HET | 2000 | 30.4 | 101 | [17,49] |
| | | HOM | 2000 | 31.5 | 71 | [20,46] |